# A population of dust-enshrouded objects orbiting the Galactic black hole

Anna Ciurlo[1]*, Randall D. Campbell[2], Mark R. Morris[1], Tuan Do[1], Andrea M. Ghez[1], Aurélien Hees[3], Breann N. Sitarski[4], Kelly Kosmo O'Neil[1], Devin S. Chu[1], Gregory D. Martinez[1], Smadar Naoz[1] & Alexander P. Stephan[1]

[1]Department of Physics and Astronomy, University of California, Los Angeles, CA, USA
[2]W. M. Keck Observatory, Waimea, HI, USA
[3]SYRTE, Observatoire de Paris, Université PSL, CNRS, Sorbonne Université, LNE, Paris, France
[4]Giant Magellan Telescope Organization, Pasadena, CA, USA
*e-mail: ciurlo@astro.ucla.edu

**The central 0.1 parsecs of the Milky Way host a supermassive black hole identified with the position of the radio and infrared source Sagittarius A\* (refs[1,2]), a cluster of young, massive stars (the S stars[3]) and various gaseous features[4,5]. Recently, two unusual objects have been found to be closely orbiting Sagittarius A\*: the so-called G sources, G1 and G2. These objects are unresolved (having a size on the order of 100 astronomical units, except at periapse where the tidal interaction with the black hole stretches them along the orbit) and they show both thermal dust emission and line emission from ionized gas[6–10]. G1 and G2 have generated attention because they appear to be tidally interacting with the supermassive Galactic black hole, possibly enhancing its accretion activity. No broad consensus has yet been reached concerning their nature: the G objects show the characteristics of gas and dust clouds but display the dynamical properties of stellar-mass objects. Here we report observations of four additional G objects, all lying within 0.04 parsecs of the black hole, and forming a class that is probably unique to this environment. The widely varying orbits derived for the six G objects demonstrate that they were commonly but separately formed.**

We used near-infrared (NIR) spectro-imaging data obtained over the past 13 years[11] at the W. M. Keck Observatory with the OSIRIS integral field spectrometer[12], coupled with laser guide star adaptive optics wave front corrections[13]. OSIRIS data-cubes have two spatial dimensions –about 3 arcsec × 2 arcsec surrounding Sgr A\* with a plate-scale of 35 mas– and one wavelength dimension –covering the Kn3 band, 2.121–2.229 μm, with a spectral resolution of $R \approx 3{,}800$. We selected 24 datacubes based on image-quality and signal-to-noise ratio; see Methods section 'Observations'. These cubes were processed through the OSIRIS pipeline[14]. We also removed the stellar continua to isolate emission features associated with interstellar gas (Methods section 'Continuum subtraction'). The reduced data-cubes were analysed with a 3D visualization tool, OsrsVol[15], that simultaneously displays all dimensions of the data-cube. This helps disentangle the many features of this crowded region, which are often superimposed in the spatial dimension but are separable in the wavelength dimension (Fig. 1).

Analysing the data with OsrsVol as well as conventional 2D and 1D tools, we identify four new compact objects in Brackett-γ line emission (Brγ; 2.1661 μm rest wavelength) that consistently appear in the data across the observed timeline. In addition to Brγ, all four objects show two [Fe III] emission lines (at 2.1457 μm and 2.2184 μm; ref.[16]).

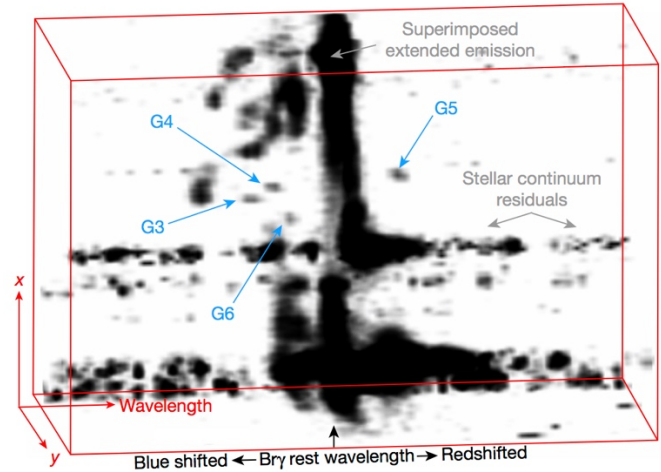

**Fig. 1 | 2006 OSIRIS data-cube visualized with OsrsVol.** The spatial dimensions ($x − y$) cover the OSIRIS field of view. The wavelength dimension is centred around Brγ (±1,500 km s$^{-1}$). G3, G4 and G6 are blueshifted while G5 is redshifted. G1 and G2 are not visible here because they have larger velocities. The extended emission in the middle is near the rest wavelength and it arises from foreground or background gas ('Superimposed extended emission'). The emission extending the full length of the wavelength axis at a few positions ('Stellar continuum residuals') is associated with continuum subtraction residuals. For this analysis, we only use sources that appear throughout the observed timeline.

The four objects show many properties in common with G1 and G2 (compact Brγ emission and coherent orbital motion) and we therefore name them G3, G4, G5 and G6. G3 was previously identified (D2[7,17]). For this work we independently identified G3 in Brγ emission, and G4, G5 and G6[a] are newly reported. We estimate that we are able to detect G objects having Brγ flux densities of at least 0.02 mJy, if they lie in a non-confused location.

Several other infrared-excess sources have been identified with L′ and K′ observations (central wavelengths of 2.2 μm and 3.8 μm, respectively[7,17], see Extended Data Fig. 1). We do not include these other sources in this work (except for G3/D2), either because they lie outside the OSIRIS field of view, or because they have not been detected in Brγ, or because they have not been consistently detected throughout the 13 years of data. We use Keck/NIRC2 L′ imaging data to investigate whether G3, G4, G5 and G6 have detectable L′ counterparts, as G1 and G2 do (Methods section 'L′ detection analysis'). No L′ counterpart was detected for G4, G5 and G6, with upper limits to the flux density of 0.4 mJy, 0.6 mJy and 0.5 mJy, respectively. G3 is detected in L′ with a dereddened flux density of 2.5 mJy, consistent with a previous report[17].

None of the G objects was detected in the K continuum. Our detection limit in the K continuum is 0.01 mJy in the OSIRIS spectra (Kn3 filter) and in K′ broad-band (2.12 μm central wavelength) a limit of 0.07 mJy was reported for G2[18] (but see[19]).

The Brγ emission is a key defining feature of the G objects because it probably results from external ionization and does not depend on the mass of a putative central object, and hence its presence is independent of the nature of the G objects (low-mass cloud or extended stellar-mass object). The compactness of such emission is what distinguishes the G objects from other presumably short-lived gas blobs that have become detached from larger-scale interstellar structures. The dust heating can be attributed to some combination of the external radiation field and an internal stellar core, if present. Therefore, the lack of detection of the G objects in the L′ band does not necessarily have implications for the existence of a stellar object embedded within the ionized external envelope.

---

[a] Recently, G6 has been independently examined[57], and interpreted as a bow shock source instead of a G object.

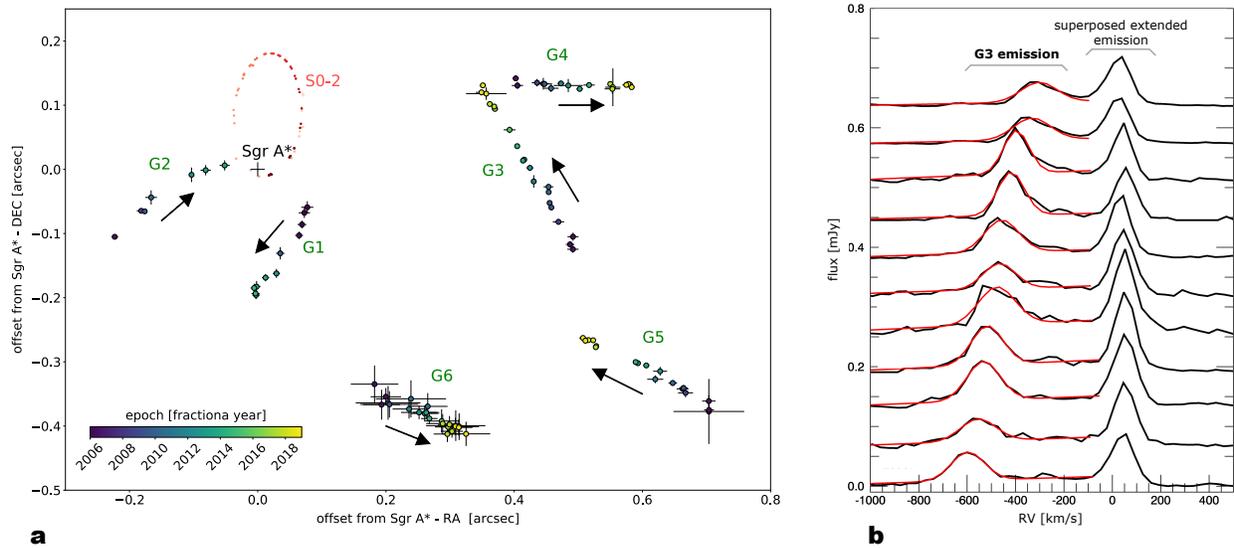

**Fig. 2 | Proper motion and spectrum of the G objects. a** Observed proper motions (with error bars showing standard deviations) of G objects and S0-2 on the plane of the sky. RA, right ascension; dec., declination. **b** G3 spectrum (black) and Gaussian fit to the G3 Brγ emission (red) in each year. There is no detected variation in the line width, but the G3 emission line blends with neighbouring features as it changes radial velocity. The large changes in the radial velocity of G3 contrast with the static extended foreground (or background) emission at the rest velocity ('Superimposed extended emission'). G objects have Brγ emission, large proper motion and radial velocity shift, and are not detected in the K continuum.

The proper motions of the G objects were determined from the Brγ centroid in the OSIRIS data (Methods sections 'Aligning OSIRIS epochs' and 'Astrometric measurements and uncertainty'). We furthermore determined the radial velocity of each object by extracting its spectrum over a 1.5-pixel-radius aperture on each data-cube and performing a Gaussian fit to the Brγ profile (Methods section 'Radial velocity measurements and uncertainty').

All G objects show large proper motions and have substantial radial velocity shifts; the radial velocity of G3 changed by about 300 km s$^{-1}$ over 13 years (see Fig. 2b and Methods section 'Radial velocity measurements and uncertainty').

Using these measurements (Extended Data Table 2), we determined the orbits of the new G objects with a Keplerian model, using a fitting algorithm[11] (Methods section 'Orbit fitting') with six orbital parameters, two parameters accounting for systematic errors in both astrometric positions and radial velocities, and one parameter accounting for correlation within the astrometric measurements. The black hole parameters (mass and Galactic Centre distance) are considered fixed[11]. The best-fit orbits are illustrated in Fig. 3 and the orbital parameters are reported in Extended Data Table 3. These fits indicate that: (1) G3, G4 and G6 have orbits with modest eccentricities ($e \approx$ 0.15, 0.3 and 0.3, respectively), while G5 has a very eccentric orbit ($e \approx$ 0.9); (2) the orbital periods range between 170 years (for G3) and 1,600 years (for G5); (3) all orbits lie on different planes, none of which contains G1 and G2 orbits or the clockwise stellar disk[20–22]; and (4) the orbits all have periods much longer than the 13 years of observations, which implies a small orbital phase coverage (~9% and ~2% in true anomaly for G3 and G5, respectively). We have run coverage tests to assess the bias attributable to the low phase coverage (Methods section 'Dependence on priors') and the results show that the obtained orbital parameters are not significantly biased (consistent with an unbiased result to within 1-sigma).

We used a Gaussian fit to the Brγ and the brightest [Fe III] line (2.2184 μm) profiles to extract fluxes. There is no noticeable flux variation for any of the four newly reported G objects in the 13 years of observations (Methods sections 'Flux calibration', 'Flux measurements' and 'Flux and FWHM summary table'). Nor can we detect any variation in the line width, given the variations in the data quality, instrumental upgrades and the emission line blending with other features.

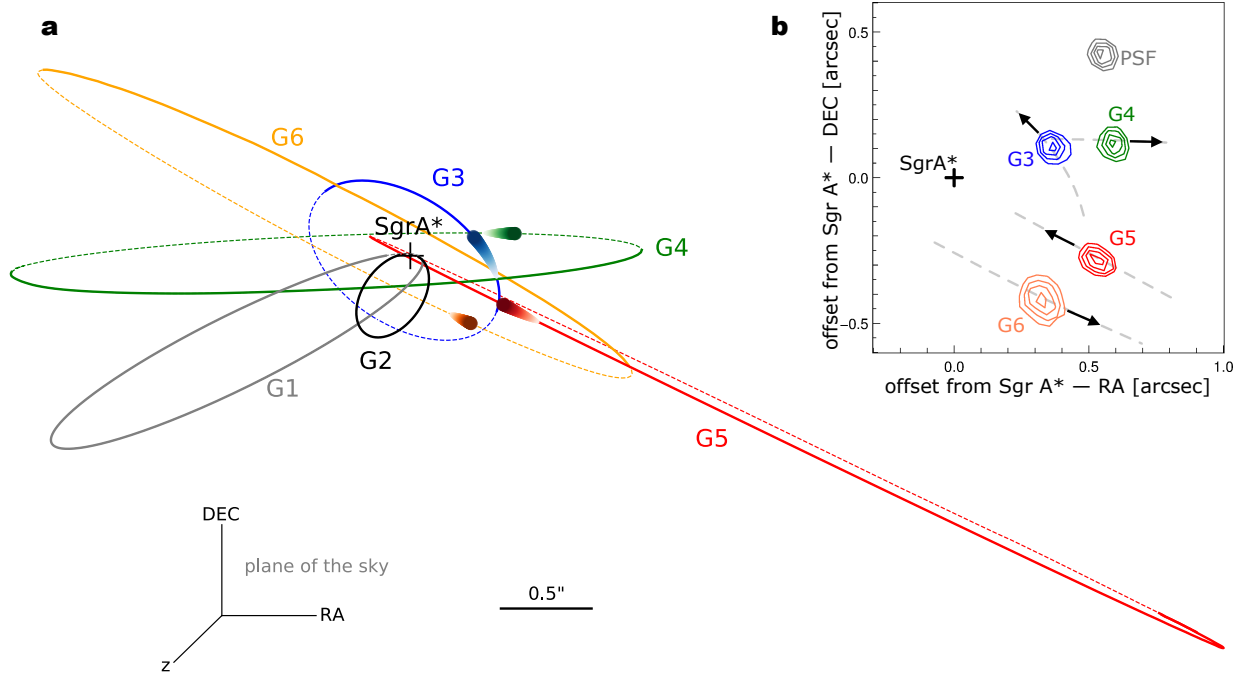

**Fig. 3 | Orbits of the G objects. a**, Orbit models in 3D: the portion of the orbit behind the plane of the sky containing Sgr A* is represented as a dashed line, while the part between the observer and Sgr A* is represented as a solid line. The thick short line indicates the time span of the observations (2006–18) and it gets darker and thicker in the direction of the object's motion. All orbits have different inclinations, eccentricities and periods. **b**, Contour plots of intensity (in Kn3 band) of the new G objects in 2018, their orbits (dashed grey lines) and the point spread function (PSF). All G objects are unresolved or only marginally resolved.

Our analysis shows that the new objects show many of the same characteristics as G1 and G2, enough to justify defining them as members of a common new class. We define the G objects to have the following characteristics: (1) presence of a distinct source of Brγ emission; (2) spatially compact emission; (3) relatively weak K-band continuum emission (such that $K' − L' \geq 4.5$); and (4) large proper motion and radial velocity shifts over time. By 'compact' we mean that they are unresolved (<0.03″) or slightly resolved (~0.05″).

These characteristics distinguish G objects clearly from normal stars. In general, the G objects seem to have very red $K − L$ colours ($K' − L' \geq 5.2$, 6 and 4.5 for G1, G2 and G3, respectively[17]), indicating that they are probably enshrouded by dust.

There are also some differences and peculiarities: G3, G4, G5 and G6 are all brighter in Brγ than G1 and G2 by about a factor of 2. G1, G2 and G3 have a clear L-band counterpart, unlike G4, G5 and G6. G3, G4, G5 and G6 show [Fe III] emission while G1 and G2 do not. G3 and G4 are unresolved, while G5 and G6 are slightly extended (Fig. 3 inset). G1 was extended after periapse[9], as G2 was before and after periapse (but reverted to being compact[10]). Despite these differences, the shared properties of the G sources warrant their aggregation into a new class with an appreciable population.

G2 was originally interpreted as an ionized gas cloud[6] and later it was argued that G1 and G2 were knots within a common orbiting filament[23]. However, this interpretation cannot apply to the new sources as they have completely different orbits. G1 and G2 have remained intact after passing through periapse and, while G2 clearly underwent tidal interaction during its periapse passage[10], its dust component has remained unresolved. This has led several authors[8,9,24–26] to suggest that there might be a stellar core shielded by an extended envelope of gas and dust. The star needs to have a relatively low mass (less than a few solar masses[20]) in order to be compatible with the weakness of the stellar continuum.

Several models (Methods section 'G-object formation scenarios') have been proposed to account for G2 in terms of an optically thick distribution of dust surrounding a star: a young, low-mass star (T Tauri star) that has retained a protoplanetary disk[25] or that generates a mass-loss envelope[26], or the merger of a binary system[8,9,24,27].

The binary merger hypothesis (in which the influence of the black hole enhances the probability of a merger through eccentricity oscillations[28]) can also account for the presence of a population of G objects by interpreting them as relatively long-lived, distended post-merger objects. Assuming the binary merger hypothesis, we have used the number of observed G objects to estimate the required binary fraction[29] in the central 0.1 pc, obtaining a lower limit of about 5% for low-mass stars (Methods section 'Binary fraction estimate'). This is compatible with the expected binary fraction[30], based on dynamical simulations[29] and taking into account the physical characteristics of the Galactic Centre. In the most likely scenario[29] for the merger hypothesis, the original binaries would have been formed in the last major star formation event at the Galactic Centre (4–6 Myr ago[22]).

Therefore, the binary merger hypothesis offers a compelling explanation for the origin of the population of G objects for several reasons: (1) it fits well with the three-body dynamics that are necessarily at play in a dense stellar environment; (2) it is compatible with the observed wide range of G-object eccentricities[27]; and (3) it fits well with the known star formation history and observed stellar population.

The random distribution of the orbital planes and the broad range of eccentricities of the G objects very closely resemble the characteristics of the orbits of the S stars, which more or less occupy the same volume. In all of the star-centred hypotheses for the G objects, the stellar object must have a relatively small mass (less than a few solar masses). At present, in the central parsec, we can directly detect stars with masses down to ~1.5 solar masses[21]. Therefore, the G objects could be offering a unique window on the low-mass, currently undetectable, part of the S-star cluster.

**Acknowledgements** We thank G. Witzel, R. Schödel and E. Becklin for providing insight and expertise. Support for this work was provided by NSF AAG grant AST-1412615, Jim and Lori Keir, the W. M. Keck Observatory Keck Visiting Scholar programme, the Gordon and Betty Moore Foundation, the Heising-Simons Foundation, and Howard and Astrid Preston. A.M.G. acknowledges support from her Lauren B. Leichtman and Arthur E. Levine Endowed Astronomy Chair. The W. M. Keck Observatory is operated as a scientific partnership among the California Institute of Technology, the University of California, and the National Aeronautics and Space Administration. The Observatory was made possible by the generous financial support of the W. M. Keck Foundation. The authors wish to recognize and acknowledge the very significant cultural role and reverence that the summit of Maunakea has always had within the indigenous Hawaiian community. We are most fortunate to have the opportunity to conduct observations from this mountain.

**Author contributions** A.C., R.D.C. and M.R.M. designed the project, and carried out the analysis and the interpretation of the results. A.M.G. supervised the observing, data acquisition and data reduction. T.D. and D.S.C. obtained and reduced the data. A.H., K.K.O'N. and B.N.S. contributed to the analysis and to the manuscript. A.C. wrote the manuscript with support from M.R.M. and R.N.C., and with contributions from all other authors. All authors provided critical feedback and helped gather data.

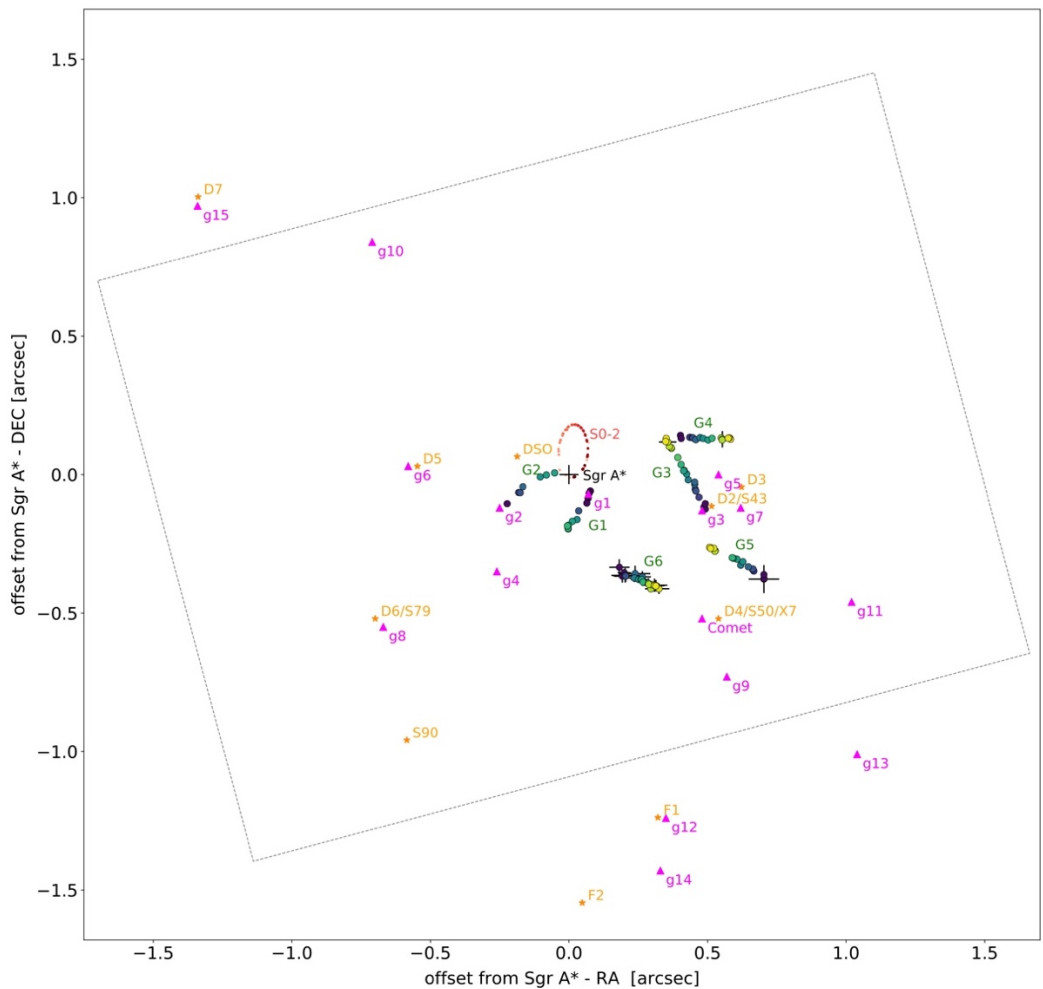

**Extended Data Fig. 1 | Infrared-excess sources and G objects.** Shown are the proper motions of the G sources (from 2006, in blue to 2018, in yellow) along with the positions of infrared-excess sources[7] in orange (data obtained with NACO at the VLT in 2005) and[17] in magenta (data obtained with NIRC2 at Keck in 2005). The red trace shows the proper motion of S0-2 (from NIRC2) as reference. The average OSIRIS field is displayed as a dotted rectangle.

## METHODS

### Observations

The observations were carried out with OSIRIS LGSAO covering 13 years, as shown in Extended Data Table 1.

| Date | Used for orbit fit | # of frames | FWHM [mas] | LSR corr [km/s] |
|---|---|---|---|---|
| 2006-06-18 | G3 G4 G5 G6 | 9 | 65 | 10 |
| 2006-06-30 | G3 G4 G5 G6 | 9 | 59 | 4 |
| 2006-07-01 | G3 - G5 G6 | 9 | 64 | 4 |
| 2008-07-25 | G3 G4 G5 G6 | 11 | 60 | -7 |
| 2009-05-05 | G3 G4 G5 - | 12 | 60 | 30 |
| 2009-05-06 | G3 G4 G5 G6 | 12 | 69 | 30 |
| 2010-05-05 | G3 G4 G5 G6 | 6 | 67 | 30 |
| 2010-05-08 | G3 - - G6 | 11 | 69 | 29 |
| 2011-07-10 | G3 G4 G5 G6 | 6 | 71 | 0 |
| 2012-07-22 | G3 G4 G5 G6 | 9 | 81 | -6 |
| 2013-05-14 | G3 G4 G5 G6 | 11 | 91 | 27 |
| 2013-07-27 | G3 - G5 G6 | 11 | 66 | -8 |
| 2014-07-03 | G3 G4 G5 G6 | 9 | 66 | 3 |
| 2015-07-21 | G3 - - G6 | 7 | 56 | -5 |
| 2017-05-17 | G3 G4 G5 G6 | 11 | 73 | 25 |
| 2017-05-18 | - G4 - G6 | 9 | 94 | 25 |
| 2017-05-19 | - G4 - G6 | 6 | 86 | 24 |
| 2017-07-19 | G3 - - G6 | 12 | 77 | -4 |
| 2017-08-14 | G3 - G5 G6 | 8 | 75 | -14 |
| 2018-04-24 | G3 - G5 G6 | 7 | 73 | 34 |
| 2018-05-23 | - G4 G5 G6 | 14 | 91 | 23 |
| 2018-07-22 | - G4 - G6 | 11 | 77 | -6 |
| 2018-07-31 | G3 G4 G5 G6 | 11 | 73 | -9 |
| 2018-08-11 | G3 G4 G5 G6 | 9 | 79 | -13 |

**Extended Data Table 1 | OSIRIS observations used for the orbital fit**
The date is reported in YYYY-MM-DD format. All frames had integration times of 900 s. The FWHM value reported is for the star S0-2. To obtain radial velocities in the local standard of rest (LSR) reference frame, each observed RV is corrected for the Earth's rotation, its motion around the Sun, and the Sun's peculiar motion with respect to the LSR.

For each epoch of observations, the OSIRIS configuration with a plate-scale of 0.035″ per lenslet was used with the Kn3 (2.121–2.229 μm) bandpass. A dither sequence with 900 s per integration using a square box pattern centred on Sgr A* with 1.0 arcsec spacing was employed to increase the field of view and to help average-out systematic instrumental features. The data were reduced using the OSIRIS data reduction package, DRP[14]. The DRP produces a wavelength-calibrated data-cube with two spatial dimensions and one spectral dimension, with the dither sequence median combined into a mosaic.

### Continuum subtraction

In order to extract the emission line of the interstellar medium we need to remove the continuum emission coming from the numerous stars in the field. To do so, we selected several spectral ranges devoid of spectral features. These spectral ranges are the same for all epochs and they are chosen to optimize the continuum estimation across the field and across the spectral band. Afterwards, we model the continuum pixel-by-pixel using a spline function. The continuum subtraction is somewhat more complex at the edges of the filter's band but this does not affect our measurements: the continuum around the emission line closest to the edge of the band that we are considering, [Fe III] 2.2184 μm, is still well modelled. We then produce new data-cubes in which the modelled continuum has been subtracted from each spectrum and use those for the rest of the analysis.

### L′ detection analysis

The Galactic Center Group has gathered L′ (at 3.8 μm) imaging data in the L′ bandpass (at 3.8 μm) with the NIRC2 imager at the W. M. Keck Observatory over several of the same epochs observed by OSIRIS and used in this study. These data were analysed to determine whether there are L′ sources coincident with the OSIRIS-detected Brγ sources via the point spread function (PSF)-fitting tool StarFinder[31]. We chose the deepest L′ epoch (2012.551[17]) to search for coincident L′ sources. No L′ counterpart was detected for G4, G5 and G6 and we perform star-planting simulations to determine an upper magnitude limit.

We used the Brγ positions of the sources and transformed them into the 2012.551 L′ coordinate system using a series of linear transformations that take into account stretching, linear offsets, and rotation. For each source, neighbouring L′ sources were subtracted out using the flux values identified with StarFinder. K′-identified sources that were not associated with the L′ sources based on proper motions were also subtracted from the analysis image assuming that they had the same magnitude and colour profiles as our flux calibration sources (S0-2, S0-12, S1-20 and S1-1[9,17]). The images were then background-subtracted and Lucy–Richardson deconvolved using the background map and model PSF generated from StarFinder. We deconvolved for 8,196 iterations and re-convolved each image with a 3-pixel full-width at half-maximum (FWHM) two-dimensional Gaussian PSF. Point sources of varying magnitude were planted in the image at the positions of G4, G5 and G6 at varying magnitude until they could no longer be detected with a modified version of StarFinder[9,18,32]. These magnitudes were then corrected for Galactic Centre extinction[33] and converted to flux densities. The L′ flux density values for G4, G5 and G6 represent upper limits, but the G5 value may still be contaminated by structured background in that region. The flux density values for G3 are consistent with previous reports[17]. All flux densities are reported in Extended Data Table 4. In all the above analyses, the single PSF model generated by StarFinder is adequate to use in this case as the off-axis positions of the candidate G sources do not experience a strong effect of the field-dependent PSF. A by-eye search for G4, G5 and G6 was performed using the L′ data coincident with the other OSIRIS epochs, but no sources were cleanly identified as being associated with the three candidate G sources. All deconvolved images in the L′-coincident epochs are shown in Extended Data Fig. 2.

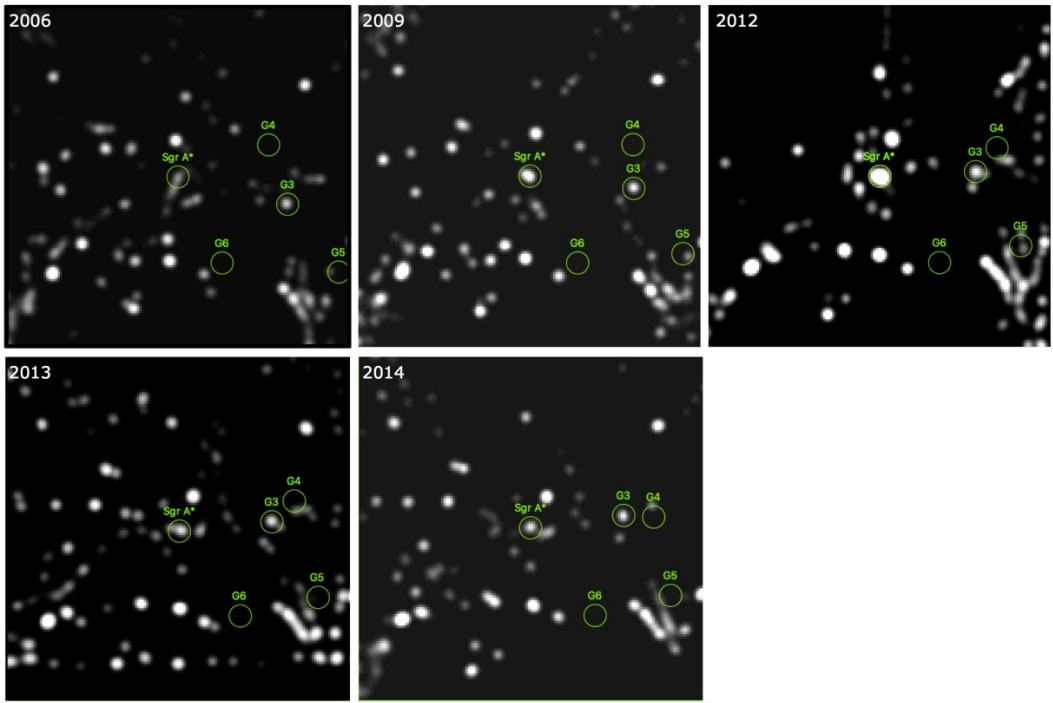

**Extended Data Fig. 2 | New G objects in the L band.** We show the deconvolved L′ images from NIRC2 for several epochs (each year is reported in the top-left corner). The green circles indicate the position of the G objects in Kn3 OSIRIS band and of Sgr A*.

## Aligning OSIRIS epochs

For our measurements and analysis, we used 24 epochs of OSIRIS observations. Each epoch consists of a mosaic constructed from frames that have been observed while dithering around the position of the star S0-2. The mosaic is obtained through a median-combine procedure applied through the OSIRIS DRP[14,34]. In order to extract the astrometry of the G objects, we shifted all mosaics into a common reference frame. To do so we measured the position of two reference stars: S0-12 and S0-14. The choice of these two specific stars was made because they are reasonably well-isolated in this crowded field, they are reasonably bright (for S0-12 $K \approx 14.3$, and for S0-14 $K \approx 13.7$; ref.[21]), and they are close to the observed G objects, thereby minimizing possible systematics in the alignment procedure due to distortion. We have accurate knowledge of the orbital motions—and thus astrometric positions—of these two stars with respect to Sgr A* from previous publications[11]. Taking into account the reference stars' motions we can put all observations in a common reference frame with Sgr A* at the centre. However, this assumes there is no significant differential distortion from epoch to epoch and that Sgr A* does not move. Given the small field of view covered by OSIRIS at this platescale, any differential distortion should be insignificant. The main source of uncertainty in this procedure comes from the centroid of the G objects. On the other hand, the position of the reference stars is very well measured because of the very high signal-to-noise. We also consider an additional systematic uncertainty on the astrometric position in the orbital fit (Methods section 'Orbit fitting').

**Astrometric measurements and uncertainty**

Analysis of the proper motion of the G objects was performed using two sets of cubes: those that had been processed to remove the continuum (Methods section 'Continuum subtraction') and those with the stellar continuum included. The G sources do not have a continuum detection in the Kn3 bandpass and thus we used the continuum-subtracted cubes to measure their positions. The positions of the G objects were measured in a median-collapsed 2D image produced by combining five spectral channels centred on the peak wavelength of the Brγ emission from each G source for each epoch. The peak-fit IDL routine was used to measure the $X$–$Y$ position in each cube. The $X$–$Y$ positions were transformed into RA–dec. coordinates relative to Sgr A* using the positions for S0-12 and S0-14 to establish the frame of reference.

S0-12 and S0-14 are stellar sources with well-established position offsets from Sgr A*, they are relatively isolated spatially, and their motion on the plane of the sky is relatively small over this time frame. The stellar positions were measured using the IDL peak-fit routine of a median-combined 2D image produced from collapsing the spectral dimension of the cube over the range of 2.133–2.158 μm (corresponding to channels 50–150). This wavelength range was chosen because it is a clean part of the spectrum that avoids emission, stellar absorption and atmospheric absorption features. The stellar point sources were mapped to a coordinate system in which Sgr A* is at rest[19,35,36]. The errors of the position measurements were estimated using a Monte Carlo method with many trials of centroid measures over variable aperture size and position. The measurements are reported in Extended Data Table 2.

**Radial velocity measurements and uncertainty**

The spectra for each G object were extracted from the continuum-subtracted data-cubes. To extract a 1D spectrum for the purpose of measuring radial velocity, the intensity of each Brγ emission feature was measured at each spectral channel of the data-cube, summing over a 1.5-pixel-radius aperture centred on the peak position of the emission feature.

The emission-line profile was analysed using a Gaussian fitting routine on the emission feature. The fits were performed on a wavelength range that isolates the feature under study as much as possible from other nearby emission features, such as the ambient gas and other G objects. The Gaussian parameter fit yields the central wavelength of the Brγ emission line, from which the radial velocity can be calculated relative to the local standard of rest.

Extended Data Fig. 3 displays the extracted spectrum and Gaussian fit for each object as it progresses over time. Changes in radial velocity over the 13-year period are evident for each G object. The velocity measurement errors were computed using the statistical errors of the Gaussian fit. In addition to the detection of Brγ emission, G3, G4, G5 and G6 display [Fe III] emission at the same Doppler shift, as shown in Extended Data Fig. 4, which displays the full Kn3 bandpass spectra from 2006-combined data sets for G3, G4, G5 and G6 (as well as G1 and G2 for comparison). G3, G4, G5 and G6 have [Fe III] detections at 2.2184 μm and 2.1457 μm (ref.[16]) while G1 and G2 only show Brγ emission. None of the G objects shows $H_2$ emission (2.1220 μm), although $H_2$ is evident in the ambient background material near zero velocity. The measurements are reported in Extended Data Table 2.

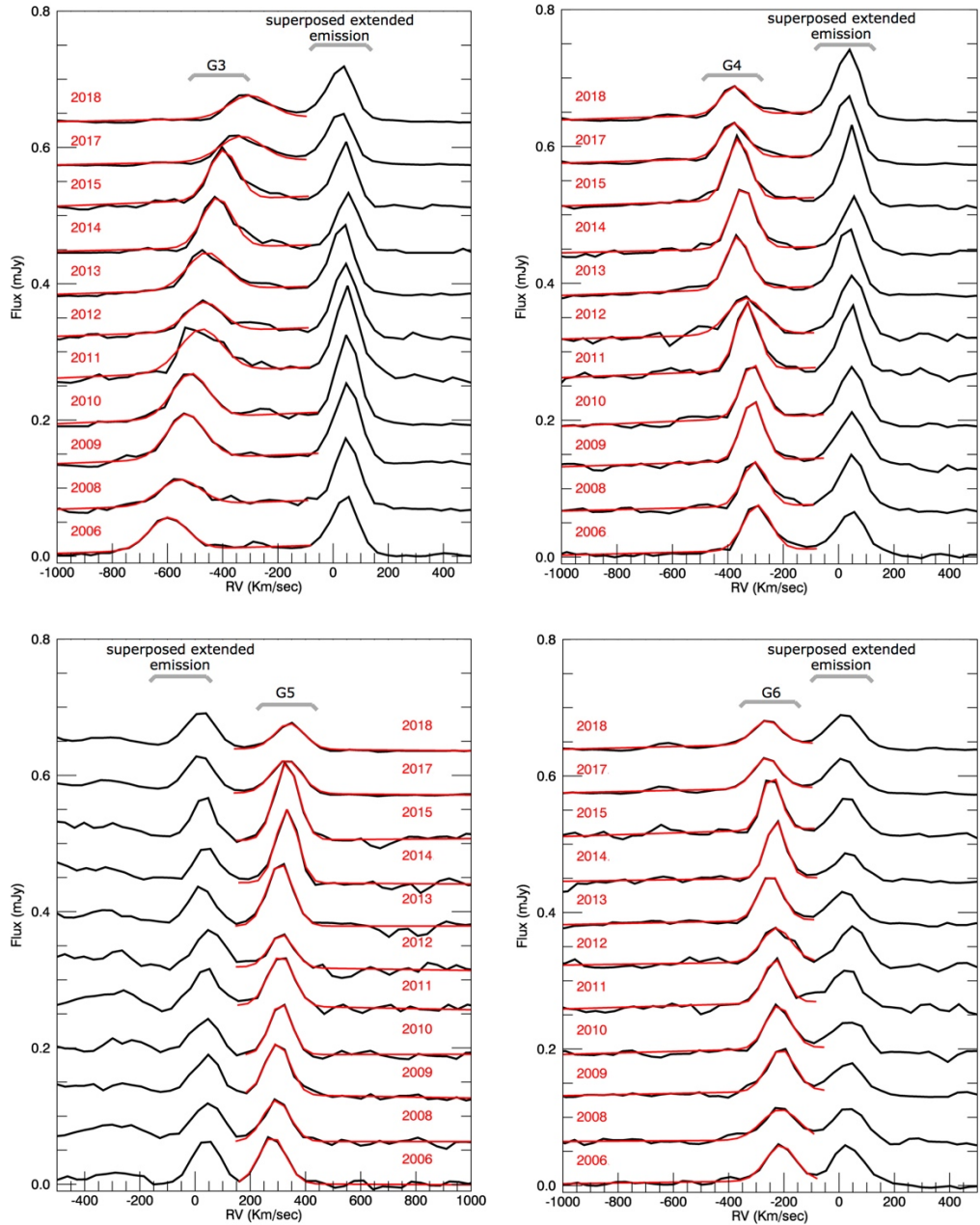

**Extended Data Fig. 3 | Spectra of the new G objects showing the Brγ emission line over time.** Top row, G3 (left) and G4 (right); bottom row, G5 (left) and G6 (right). The spectra are extracted epoch by epoch (black). The Gaussian fit of the Brγ emission line (red) is superimposed. There is no significantly detected variation (all values are compatible within 1-sigma) in the linewidth for any of the objects. The data quality varies and the emission of the objects blends with neighboring features as it changes radial velocity (RV) and position: this gives sometimes the impression of a broadening of the line which is not real. The emission line at the rest velocity is part of the extended emission present across the field and does not change with time.

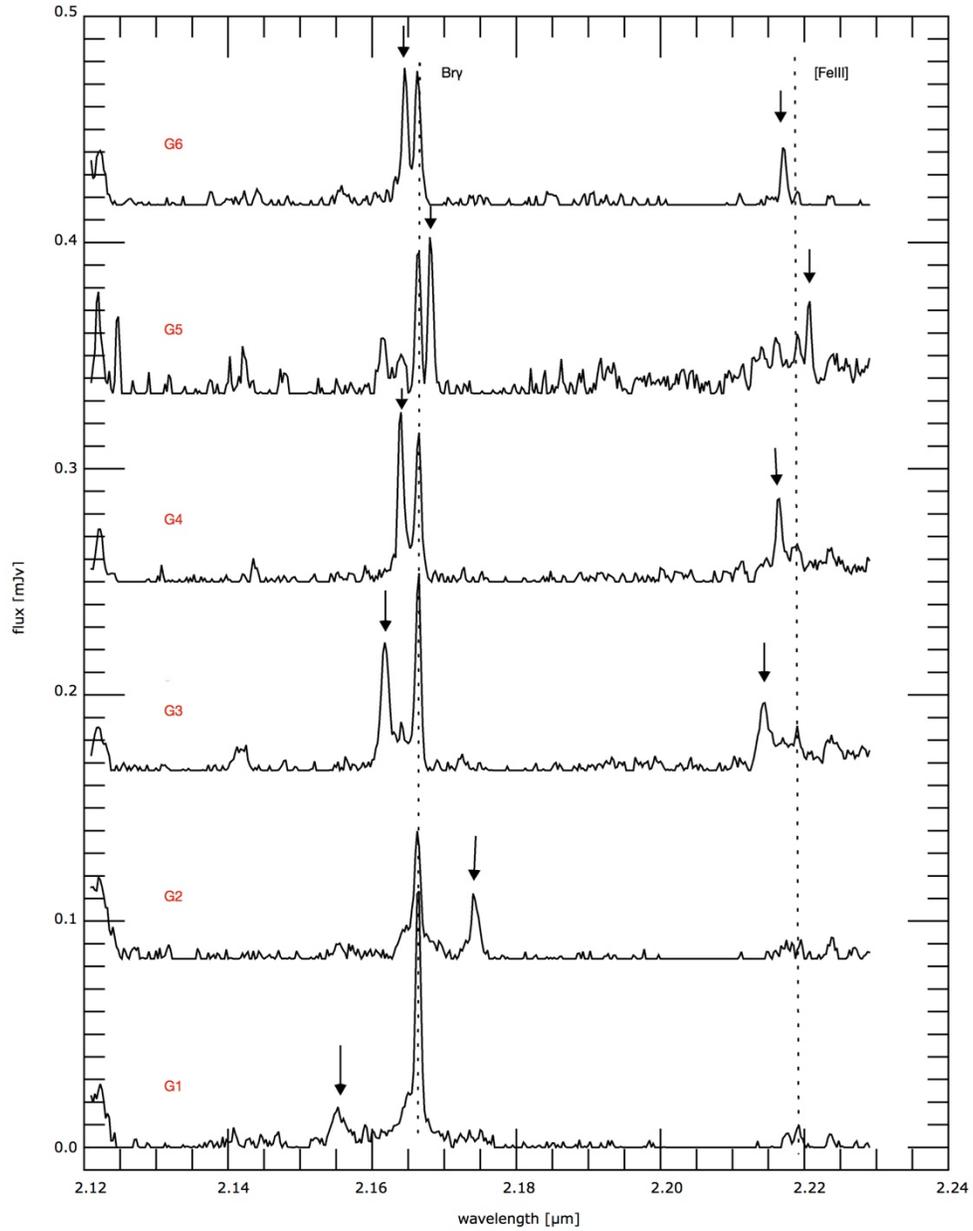

**Extended Data Fig. 4 | Spectra of the G objects showing Brγ and [Fe III] emission lines.** The Kn3 spectra of G objects G6–G1 are extracted over an aperture of 1.5-pixel radius from the 2006 combined data set. The dotted lines show the rest-frame velocity of the Brγ and [Fe III] emission lines. G3, G4, G5 and G6 show both Brγ and [Fe III] emission moving at the same velocity (Doppler-shifted emission indicated by the arrows), while G1 and G2 only show Brγ emission.

| | epoch | RA | err | DEC | err | RV | err | | epoch | RA | err | DEC | err | RV | err |
|---|---|---|---|---|---|---|---|---|---|---|---|---|---|---|---|
| | 53904.4 | 0.487 | 0.007 | -0.117 | 0.005 | -594 | 2 | | 53904.4 | 0.402 | 0.005 | 0.142 | 0.004 | -293 | 1 |
| | 53916.3 | 0.492 | 0.009 | -0.125 | 0.006 | -600 | 2 | | 53916.3 | 0.405 | 0.009 | 0.131 | 0.006 | -291 | 2 |
| | 53917.3 | 0.492 | 0.009 | -0.105 | 0.006 | -592 | 2 | | 54672.3 | 0.436 | 0.008 | 0.135 | 0.006 | -304 | 1 |
| | 54672.3 | 0.469 | 0.009 | -0.082 | 0.005 | -555 | 3 | | 54956.5 | 0.445 | 0.009 | 0.133 | 0.009 | -310 | 1 |
| | 54956.5 | 0.455 | 0.004 | -0.052 | 0.001 | -536 | 2 | | 54957.5 | 0.447 | 0.012 | 0.133 | 0.004 | -313 | 1 |
| | 54957.5 | 0.458 | 0.004 | -0.059 | 0.003 | -531 | 2 | | 55321.5 | 0.458 | 0.012 | 0.126 | 0.006 | -314 | 1 |
| | 55321.5 | 0.454 | 0.003 | -0.035 | 0.003 | -518 | 1 | | 55752.3 | 0.473 | 0.005 | 0.134 | 0.004 | -330 | 1 |
| | 55324.5 | 0.454 | 0.008 | -0.027 | 0.002 | -512 | 2 | | 56130.3 | 0.484 | 0.023 | 0.131 | 0.010 | -334 | 2 |
| G3 | 55752.3 | 0.431 | 0.004 | -0.019 | 0.010 | -478 | 3 | G4 | 56426.5 | 0.502 | 0.005 | 0.125 | 0.002 | -345 | 1 |
| | 56130.3 | 0.425 | 0.005 | 0.002 | 0.004 | -464 | 3 | | 56841.3 | 0.517 | 0.009 | 0.131 | 0.005 | -346 | 1 |
| | 56426.5 | 0.416 | 0.002 | 0.015 | 0.001 | -450 | 2 | | 57890.4 | 0.550 | 0.004 | 0.133 | 0.003 | -360 | 1 |
| | 56500.3 | 0.414 | 0.004 | 0.014 | 0.002 | -438 | 2 | | 57891.5 | 0.554 | 0.012 | 0.128 | 0.029 | -360 | 1 |
| | 56841.3 | 0.405 | 0.002 | 0.036 | 0.001 | -419 | 1 | | 57892.5 | 0.553 | 0.012 | 0.125 | 0.008 | -360 | 1 |
| | 57224.3 | 0.393 | 0.009 | 0.062 | 0.005 | -398 | 1 | | 58261.4 | 0.581 | 0.004 | 0.132 | 0.003 | -361 | 1 |
| | 57890.4 | 0.370 | 0.005 | 0.094 | 0.002 | -324 | 3 | | 58321.3 | 0.583 | 0.005 | 0.128 | 0.002 | -368 | 2 |
| | 57953.3 | 0.369 | 0.003 | 0.098 | 0.002 | -312 | 2 | | 58330.3 | 0.578 | 0.005 | 0.133 | 0.004 | -363 | 2 |
| | 57979.2 | 0.362 | 0.005 | 0.102 | 0.003 | -318 | 2 | | 58341.3 | 0.575 | 0.006 | 0.132 | 0.003 | -365 | 1 |
| | 58232.5 | 0.356 | 0.032 | 0.118 | 0.010 | -298 | 3 | | epoch | RA | err | DEC | err | RV | err |
| | 58330.3 | 0.351 | 0.005 | 0.131 | 0.003 | -288 | 2 | | 53904.4 | 0.19 | 0.03 | -0.37 | 0.02 | -210 | 2 |
| | 58341.3 | 0.349 | 0.009 | 0.120 | 0.005 | -300 | 2 | | 53916.3 | 0.18 | 0.04 | -0.33 | 0.03 | -207 | 2 |
| | epoch | RA] | err | DEC | err | RV | err | | 53917.3 | 0.20 | 0.02 | -0.35 | 0.01 | -209 | 1 |
| | 53904.4 | 0.70 | 0.06 | -0.38 | 0.05 | 267 | 2 | | 54672.3 | 0.20 | 0.05 | -0.36 | 0.03 | -210 | 2 |
| | 53916.3 | 0.70 | 0.01 | -0.37 | 0.01 | 281 | 2 | | 54957.5 | 0.21 | 0.05 | -0.37 | 0.02 | -210 | 1 |
| | 53917.3 | 0.70 | 0.01 | -0.36 | 0.01 | 276 | 1 | | 55321.5 | 0.24 | 0.05 | -0.36 | 0.03 | -212 | 2 |
| | 54672.3 | 0.67 | 0.01 | -0.35 | 0.01 | 296 | 2 | | 55324.5 | 0.27 | 0.03 | -0.37 | 0.02 | -225 | 1 |
| | 54956.5 | 0.66 | 0.01 | -0.34 | 0.00 | 297 | 1 | | 55752.3 | 0.24 | 0.03 | -0.37 | 0.01 | -232 | 2 |
| | 54957.5 | 0.66 | 0.01 | -0.34 | 0.00 | 304 | 1 | | 56130.3 | 0.25 | 0.03 | -0.38 | 0.01 | -227 | 2 |
| | 55321.5 | 0.65 | 0.01 | -0.33 | 0.00 | 311 | 1 | | 56426.5 | 0.26 | 0.02 | -0.38 | 0.01 | -230 | 1 |
| | 55752.3 | 0.62 | 0.01 | -0.33 | 0.01 | 311 | 1 | | 56500.3 | 0.26 | 0.02 | -0.38 | 0.01 | -230 | 1 |
| G5 | 56130.3 | 0.63 | 0.01 | -0.31 | 0.01 | 314 | 2 | G6 | 56841.3 | 0.27 | 0.01 | -0.39 | 0.01 | -229 | 1 |
| | 56426.5 | 0.61 | 0.00 | -0.31 | 0.00 | 327 | 0 | | 57224.3 | 0.29 | 0.03 | -0.39 | 0.02 | -244 | 1 |
| | 56500.3 | 0.59 | 0.00 | -0.30 | 0.00 | 329 | 1 | | 57890.4 | 0.30 | 0.02 | -0.40 | 0.01 | -238 | 1 |
| | 56841.3 | 0.59 | 0.00 | -0.30 | 0.00 | 331 | 0 | | 57891.5 | 0.30 | 0.02 | -0.41 | 0.01 | -237 | 1 |
| | 57890.4 | 0.53 | 0.00 | -0.28 | 0.00 | 352 | 0 | | 57892.5 | 0.30 | 0.02 | -0.41 | 0.01 | -240 | 2 |
| | 57979.2 | 0.53 | 0.00 | -0.28 | 0.00 | 354 | 1 | | 57953.3 | 0.30 | 0.02 | -0.40 | 0.01 | -245 | 1 |
| | 58232.5 | 0.52 | 0.01 | -0.27 | 0.00 | 360 | 1 | | 57979.2 | 0.29 | 0.03 | -0.40 | 0.01 | -240 | 1 |
| | 58261.4 | 0.52 | 0.00 | -0.27 | 0.00 | 358 | 1 | | 58232.5 | 0.31 | 0.05 | -0.40 | 0.02 | -253 | 2 |
| | 58330.3 | 0.51 | 0.01 | -0.26 | 0.00 | 363 | 2 | | 58261.4 | 0.32 | 0.04 | -0.41 | 0.02 | -244 | 1 |
| | 58341.3 | 0.51 | 0.00 | -0.27 | 0.00 | 358 | 2 | | 58321.3 | 0.31 | 0.03 | -0.40 | 0.02 | -237 | 1 |
| | | | | | | | | | 58330.3 | 0.31 | 0.02 | -0.40 | 0.01 | -241 | 1 |
| | | | | | | | | | 58341.3 | 0.31 | 0.02 | -0.40 | 0.01 | -244 | 1 |

**Extended Data Table 2 | Measured values of positions and radial velocities**
Data are shown for G3, G4, G5 and G6. The epochs are reposted as modified Julian date (MJD). The positions (RA and dec.) are offsets from Sgr A* in arcseconds. The position uncertainties are the standard deviation obtained through a Monte Carlo method. The radial velocities (RV) are in km s$^{-1}$ and have been corrected for the local standard of reference. The reported radial velocity uncertainties (err.) are purely statistical (1  of the line fit). There is an additional systematic uncertainty that we fold into the orbit fit.

**Orbit fitting**

The astrometric and radial velocity measurements (Extended Data Table 2) are combined in a global orbit fit. The software used for orbit fitting has been previously used for the detection of the relativistic redshift on S0-2[11]. The orbital modelling assumes Keplerian motion parameterized by the six following orbital elements[2]: the period ($P$), the time of closest approach ($T_0$), the eccentricity ($e$), the inclination ($i$), the argument of periastron ($\omega$) and the longitude of the ascending node ($\Omega$). The G objects do not have enough orbital coverage and information to constrain the parameters related to the central mass (the mass of the black hole, the distance to our Galactic Centre $R_0$, the position and velocity of the central mass). Therefore, we fixed the values of the black hole mass and $R_0$ to the ones obtained from S0-2's measurements[11], that is, to $M = 3.964 \times 10^6 M_\odot$ (where $M_\odot$ is the solar mass) and $R_0 = 7.971$ kpc.

Our orbital fits are performed using Bayesian inference with a MultiNest sampler[37,38]. The radial velocity measurements are assumed to be independent and normally distributed. To take into account possible systematics at the level of the orbital fit, we use a likelihood that includes a systematic uncertainty ($\sigma_{RV}$) for the radial velocities. In summary, the radial velocity (RV) measurements are assumed to be distributed following:

$$RV_i \sim \mathcal{N}[RV(t_i), \sigma_{RV_i}^2 + \sigma_{RV}^2]$$

where $RV_i(t)$ are the predicted radial velocity values, $\sigma_{RV_i}$ are the measurement uncertainties and where $x \sim \mathcal{N}[\mu, \sigma^2]$ denotes that $x$ is normally distributed around $\mu$ with a variance of $\sigma^2$. On the other hand, the astrometric measurements are assumed to be correlated, that is, the likelihood is assumed to be a multivariate normal distribution characterized by a covariance matrix. In addition, to take into account possible systematics at the level of the orbital fit, we also include an additional parameter: a systematic uncertainty for the astrometry, $\sigma_{astro}$. The astrometric measurements are therefore assumed to be distributed as:

$$\mathbf{x} \sim \mathcal{N}[x(\mathbf{t}_{astro}), \Sigma_x] \text{ and } \mathbf{y} \sim \mathcal{N}[y(\mathbf{t}_{astro}), \Sigma_y],$$

where $x(\mathbf{t}_{astro})$ and $y(\mathbf{t}_{astro})$ are the predicted astrometric values, $\Sigma_x$ and $\Sigma_y$ are the covariance matrices and where $\mathbf{x} \sim \mathcal{N}[\boldsymbol{\mu}, \Sigma]$ denotes that the vector $\mathbf{x}$ is normally distributed around the vector $\boldsymbol{\mu}$ with a covariance matrix of $\Sigma$. We model the covariance matrices[11] by:

$$[\Sigma_x]_{ij} = [\rho]_{ij} \sqrt{\sigma_{x_i}^2 + \sigma_{astro}^2} \sqrt{\sigma_{x_j}^2 + \sigma_{astro}^2} \text{ and } [\Sigma_y]_{ij} = [\rho]_{ij} \sqrt{\sigma_{y_i}^2 + \sigma_{astro}^2} \sqrt{\sigma_{y_j}^2 + \sigma_{astro}^2}$$

where $\sigma_{astro}$ is the systematic uncertainty and $\rho$ is the correlation matrix that characterizes the correlation of the measurement errors. This correlation matrix is given by[11]:

$$[\rho]_{ij} = (1-c)\delta_{ij} + c \cdot e^{-|d_{ij}|/\Lambda}$$

where $\delta_{ij}$ is kronecker delta and $d_{ij}$ is the 2D projected distance between point $i$ and point $j$:

$$d_{ij} = \sqrt{(x_i - x_j)^2 + (y_i - y_j)^2}$$

Here $\Lambda$ is a correlation length scale that typically takes the value of half the diffraction limit of the detector[11], and is fixed here at a value of 35 mas; $c$ is a mixing parameter that is fitted simultaneously with all parameters and that characterizes the strength of the correlation. Corner plots of the best fit are shown in Extended Data Fig. 5 and the best fit parameters are reported in Extended Data Table 3.

In addition, we use uniform priors on all fitted parameters. We show in the next Methods section that this does not bias our estimates.

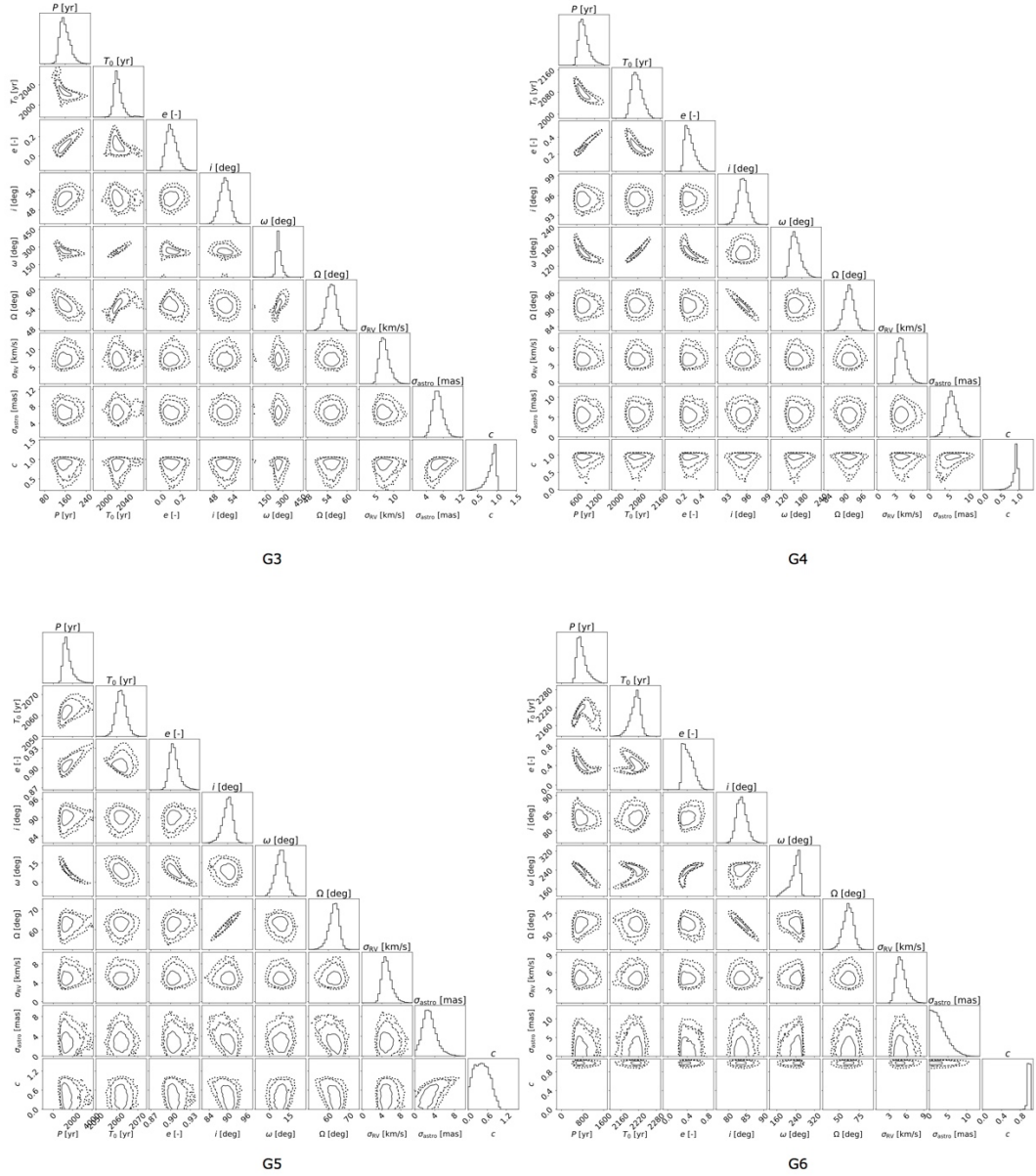

**Extended Data Fig. 5 | Corner plots for orbit fitting of G3, G4, G5 and G6.** See Methods section 'Orbit fitting' for details of the parameters displayed.

|    | P [yr] | T0 [yr] | e | i [deg] | ω | Ω | $\sigma_{astro}$ | $\sigma_{RV}$ | c |
|----|--------|---------|---|---------|---|---|------------------|---------------|---|
| G3 | $156^{+22}_{-16}$ | $2024^{+9}_{-6}$ | $0.11^{+0.06}_{-0.05}$ | $52^{+1.5}_{-1.6}$ | $261^{+23}_{-16}$ | $55^{+1.5}_{-1.6}$ | $6.7^{+1.3}_{-1.1}$ | $7.2^{+1.6}_{-1.1}$ | $0.89^{+0.08}_{-0.15}$ |
| G4 | $815^{+190}_{-126}$ | $2075^{+21}_{-16}$ | $0.28^{+0.07}_{-0.05}$ | $95.4^{+1.7}_{-0.7}$ | $160^{+18}_{-13}$ | $92^{+1.7}_{-1.8}$ | $5.5^{+1.5}_{-1.3}$ | $4^{+0.9}_{-0.7}$ | $0.96^{+0.03}_{-0.1}$ |
| G5 | $1392^{+555}_{-315}$ | $2062^{+2}_{-2}$ | $0.9^{+0.01}_{-0.01}$ | $90.1^{+1.4}_{-1.7}$ | $8.8^{+3.7}_{-4.1}$ | $62.9^{+2.1}_{-2.6}$ | $2.8^{+1.5}_{-1.2}$ | $4.9^{+1.1}_{-0.8}$ | $0.42^{+0.27}_{-0.25}$ |
| G6 | $732^{+224}_{-136}$ | $2211^{+13}_{-19}$ | $0.41^{+0.13}_{-0.09}$ | $83.6^{+1.5}_{-1.3}$ | $242^{+13.8}_{-26.7}$ | $61.7^{+5.8}_{-6.4}$ | $2.4^{+2.6}_{-1.7}$ | $5^{+0.9}_{-0.7}$ | $0.99^{+0.01}_{-0.02}$ |

**Extended Data Table 3 | Best fit orbital parameters**
We show the median and central 68% confidence interval of the best-fit orbital parameters for G3, G4, G5 and G6. See Methods section 'Orbit fitting' for nomenclature.

**Dependence on priors**

To estimate the orbits of the G objects, we use uniform priors on all eight fitted parameters (six orbital parameters and two systematic uncertainty parameters). While uniform priors are commonly assumed in orbit fitting, this choice has been shown to cause potential biases in estimated parameters when orbital periods are much longer than the time baseline of observations[39,40]. To assess the impact of our fitting procedure in this context, we ran simulations to assess possible biases in the estimated parameters and to test the accuracy of confidence intervals obtained in our analysis[40]. We generated 100 mock data sets with simulated measurements at epochs corresponding to our observations. The simulated measurements were randomly drawn from a normal distribution about an assumed 'true' value, with a dispersion equal to the true measurement error at that epoch. We fit each of these 100 mock data sets with the same orbit fitting procedure as described above. The bias on each fitted parameter is computed from the difference between the estimated parameter value and the input parameter value, normalized by the $1\sigma$ confidence interval on the corresponding parameter. For all eight fitted parameters, the distribution of bias values is centred around zero for G3, G4, G5 and G6 within the 68% confidence interval, indicating non-biased parameter estimates.

In addition, we evaluate statistical efficiency to demonstrate that the confidence intervals used in this analysis are well-defined and have close to exact coverage. According to the classical definition of a confidence interval, $1\sigma$ confidence intervals inferred from each orbit fit should cover the 'true' value (from the simulated data) 68% of the time. In other words, given 100 randomly drawn simulated data sets, a 68% confidence interval requires that about 68 out of 100 fits produce a confidence interval that covers the true value[41]. However, effective coverage (defined as the experimentally determined percentage of data sets in which the inferred confidence interval covers the true value) is rarely exact. Statistical efficiency, defined as the ratio of effective coverage to stated or expected coverage (for example, 68% for a $1\sigma$ confidence interval), is thus a powerful performance diagnostic that can be used to investigate the accuracy of calculated confidence intervals[40]. By definition, a statistical efficiency of one indicates exact coverage. The statistical efficiencies for all parameters for G3, G4 and G6 are consistent with one. For G5, the period is slightly under-covered with a statistical efficiency of 0.81 ± 0.09, indicating that the inferred confidence interval on G5's period is slightly underestimated. The statistical efficiencies for all other parameters for G5 are consistent with one. This analysis indicates that, in general, confidence intervals calculated in this work provide robust estimates of the statistical uncertainty.

**Flux calibration**

For this project we perform the absolute flux calibration of OSIRIS data. To do so, we need to apply aperture photometry to isolate sources of known magnitude. Even though many of the stellar sources are well-known, the Galactic Centre is a very crowded environment: no source is truly isolated and the combined background of underlying sources is challenging to determine. To measure the flux of stars on the field we would need to use a very small aperture radius. However, the PSF cannot be easily modelled, since the observations are taken through adaptive optics. Moreover, the OSIRIS field of view is very small, making an accurate empirical knowledge of the PSF impossible.

Instead we use observations of standard A stars, obtained the same night as the Galactic Centre observations. We used: HD 155379, HD 195500 and HD 146606 with 2MASS *K* magnitudes of 6.52, 7.19 and 7.04 respectively. These stars are chosen to be at around the same airmass as the science targets and their observations are taken as close in time as possible to the science observations (within a few hours). These are well-known, bright and isolated sources for which we can use aperture photometry over a very large radius that encompasses almost all of the source. In this way we can gather close to 100% of the flux and avoid problems related to the PSF shape.

The A-star frames are obtained by dithering around the star's position and are treated with the standard calibration procedure to remove atmospheric effects. Here, for each epoch, we use all available frames independently to measure the counts-to-Jy conversion factor and use their dispersion to estimate the corresponding uncertainty. Both the science mosaic and the A-star frames are calibrated in the standard way of the group. For each epoch, for each frame, we perform a 2D Gaussian fit to get the centroid of the source and an estimate of the Gaussian $\sigma$. We extract the A-star flux ($F_{ap}$) within a ~12-pixel aperture radius, which is ~6 times the $\sigma$ of the 2D Gaussian fit (that is, that encompass ~100% of the stars' flux). We subtract the sky background through an annulus 1 pixel larger than the aperture size and of 1-pixel thickness ($F_{an}$). We use the known magnitude of the star (from the 2MASS catalogue) to compute its expected flux in the Kn3 band ($F_{th}$) using Vega as zero-point. The conversion factor is computed as follows:

$$CF = \frac{F_{th}}{F_{ap} - F_{an}} \frac{1}{df}$$

where $df$ is the width of the spectral channel in Hz. The same process is repeated for all frames within one given epoch and the median is adopted as the value for that epoch and the dispersion as the uncertainty. We checked other potential sources of error, such as imprecise pointing on the centre of the star, but we always obtained uncertainties not several orders of magnitude smaller than the one coming from the dispersion.

The disadvantage of not using sources within the science field for calibration is that there could be variations of the fraction of photons reaching the detector surface between the science target and calibrator observations—for example, because of variations in the extinction due to passing clouds at the telescope site. However, the variation in extinction due to clouds is usually less than 0.5 mag and should have an impact smaller than the final calibration error. Indeed, the final calibration factor does not vary much from night to night or even year to year. The most dramatic variations are related to instrumental hardware upgrades. Therefore, we have chosen to divide the OSIRIS instrument timeline into three parts[14]: (1) 2006–12 before the grating upgrade; (2) 2012–16 before the spectrograph upgrade; and (3) from 2017 on.

For each of these periods, we consider the median of the conversion factors as the final value and the dispersion of the measurements as its uncertainty. This way we obtain three calibration factors with an error of about 10%. We also compare the conversion factor obtained with the A stars to the one obtained using multiple stars on the field. In the case of the field stars the values are very sensitive to the applied correction to the aperture flux, and the conversion factor therefore varies more dramatically (even within close epochs) than in the case of the A stars (Extended Data Fig. 6). Therefore, we can affirm that the flux calibration obtained through standard stars is more robust.

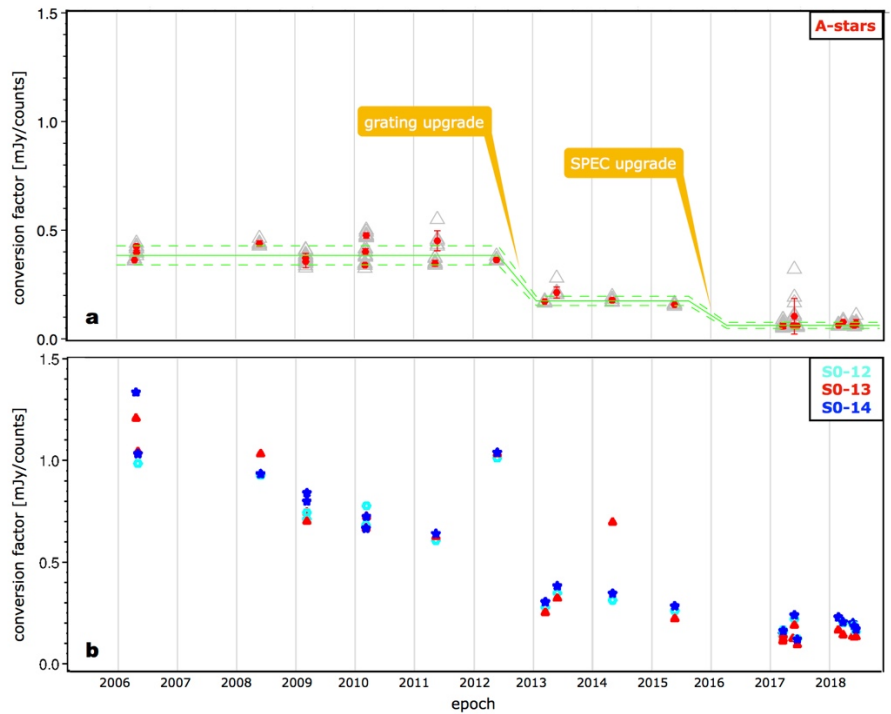

**Extended Data Fig. 6 | Unit conversion factors as a function of time. a**, Factor obtained using calibration A stars (single frames in grey triangles, median for each epoch in red dots, dispersion used as error bar). **b**, Factor obtained using stars in the science field. The dispersion when using stars in the science field is much larger. We use the A stars, for which most of the variation corresponds to hardware changes in the instrument. We use the median value for each instrument period (green solid line) and use the dispersion as an

**Flux measurements**

In order to maximize the signal-to-noise ratio, we measured the flux on data-cubes combined year by year (hence 1 cube per year). Multiple data sets were combined for each year using all available epochs to enhance the signal-to-noise ratio in the emission lines, resulting in 11 data-cubes for as many years between 2006 and 2018 (except for 2007 and 2016 where, respectively, the image quality was too poor and no Kn3 data cube was obtained).

The Brγ line fluxes of the G objects are obtained for each combined data-cube by extracting its spectrum and performing a Gaussian fit to the emission line. Flux measurements were derived from each line profile using an equivalent width method. The equivalent width was computed from the Gaussian fit parameters of the emission features from Brγ and [Fe III]. A conversion from measured flux to W m$^{-2}$ was established using the flux calibration performed for each epoch from observations of A standards (Methods section 'Flux calibration'). Note that the absolute flux calibration can have relatively high errors in AO systems where the image quality and encircled energy in the data collection can change substantially on short timescales and from night to night. The measure fluxes are dereddened[33]. The measurements are reported in the following Methods section and in Extended Data Table 4.

**Flux and FWHM summary table**

The measured flux densities for all objects are reported in Extended Data Table 4, along with measurements of the spatial and spectral width. We do not detect any continuum in the Kn3 band in any of the G sources (we find a detection limit of 0.01 mJy). However, G2 detection in K-broadband imaging data has been claimed[19], finding a dereddened flux of about 0.25 mJy in $K_s$ (2.18 μm central wavelength), which compares to a detection limit of 0.07 mJy in K' band[18] (2.12 μm central wavelength).

|    | NIRC2 | OSIRIS | | | | | |
|----|-------|--------|--|--|--|--|--|
|    |       | flux density | | FWHM | | S/N | |
|    | L' [mJy] | Brγ [mJy] | [FeIII] [mJy] | spectral [km/s] | spatial ['] | Brγ | [FeIII] |
| G3 | 2.5±0.5 | 1.16±0.15 | 0.59±0.11 | 156±5 | 0.10±0.02 | 24 | 14 |
| G4 | <0.38±0.08 | 1.10±0.14 | 0.49±0.09 | 117±2 | 0.10±0.02 | 34 | 16 |
| G5 | <0.57±0.11 | 0.99±0.13 | 0.44±0.08 | 110±2 | 0.12±0.02 | 17 | 16 |
| G6 | <0.54±0.10 | 0.82±0.12 | 0.37±0.07 | 127±5 | 0.13±0.02 | 28 | 19 |
| G1 | 0.6±0.05 | 0.48±0.15 | <0.027±0.005 | 293±20 | 0.10±0.04 | 8 | - |
| G2 | 2.12±0.15 | 0.65±0.19 | <0.027±0.003 | 174±14 | 0.08±0.04 | 13 | - |

**Extended Data Table 4 | Emission from the G objects**

We report properties of this emission: flux densities, spectral and spatial widths (as FWHM), and signal-to-noise ratios (*S/N*). The L' flux densities come from NIRC2 2012 measurements (G1 is brighter in earlier epochs[17]). The total flux density for Brγ and [Fe III] (2.2184 μm) come from OSIRIS (average of all observing). For comparison, we report G1 and G2 measurements from 2006 where all G objects are detectable in OSIRIS data. All fluxes are dereddened[33]. The spectral and spatial FWHM (an average of the *x*- and *y*-FWHM) are measured for Brγ These values are not corrected for instrumental line width and PSF size (respectively ~100 km s$^{-1}$ and ~75 mas). The *S/N* of the other [Fe III] line has a value of ~6 for all objects.

**G-object formation scenarios**

Although many hypotheses have been proposed to explain the origin of G1 and G2, the principal debate has centred on whether they are compact, dusty gas clouds or gaseous features anchored on stellar cores. G1 and G2 were first interpreted as purely gas and dust clouds[6]. However, G1 and G2 have remained intact after passing through periapse, which has led some[8,24–26] to argue that they must have a stellar core shielded by an extended opaque envelope of gas and dust.

Given the absence of photospheric emission, the original G2 hypothesis[6] interpreted it as an ionized gas cloud of 3 Earth masses. Since its discovery, the gas has been tidally interacting with the black hole. It was argued[10,23] that G1 and G2 are knots of gas and dust that have formed within a common orbiting filament. Indeed, their orbits are similar, but substantially different[9]. A drag force has been invoked to explain this difference[10]. However, the common filament interpretation cannot apply to the new sources we present here because of their very different location and orbit.

Given the strong tidal forces near the black hole, and the high flux of ultraviolet radiation in this region, compact gas clouds would supposedly be transient phenomena, unless they could be stably confined by a high external pressure[42]. Otherwise, they would need to be continuously produced in order to account for the sizable population we observe. The region is rich in gaseous interstellar medium structures, including the Epsilon source[43] (a nearby feature immediately west of the field), the Minispiral[44] and the Circumnuclear Disk[45]. It is possible that small pieces of these larger structures get detached and stay in the region for a few decades before getting destroyed, but it is not clear that such gas blobs would be as compact as the observed G sources.

The alternative hypothesis is that the G objects host a star. While G2 is tidally interacting during its closest approach to Sgr A*[10], the dust component of G2 has remained unresolved. The emitting gas is unbound at closest approach[10], but that is not inconsistent with the existence of a stellar mass keeping the dust emission compact[8]. Several models have been proposed to account for G2 in terms of an optically thick distribution of dust surrounding a star: a young, low-mass star (T Tauri star) that has retained a protoplanetary disk[25] (scenario 1) or that generates a mass-loss envelope[26] (scenario 2); or, the merger of a binary system[9,24,27] (scenario 3).

In the first scenario, G2 could be a young star that has retained a protoplanetary disk and that this star was scattered inwards from the massive cluster of young stars distributed on larger scales[36]. Stars having protoplanetary disks are common in young clusters, but it is unclear whether such disks would survive the abrupt scattering event needed to transfer the protostars onto such tight orbits around the black hole. Furthermore, protoplanetary disks do not last very long except under the most benign conditions (up to 5–7 Myr; ref.[46]), therefore a population of such objects in the particularly hostile Galactic Centre environment is not obviously compatible with the timescale of the last star formation event (4–6 Myr ago[22]). Therefore, the protostellar disk hypothesis might be strongly constrained as an explanation for the common origin of these objects unless star formation is continuous at the Galactic Centre, as some have argued[47,48]. This matter is still under debate, but any demonstration that a substantial number of protoplanetary disks have survived in the central 0.05 pc of the Galactic Centre would have important implications for our understanding of star formation in this region.

In the second scenario, G2 was proposed to be the product of the mass-losing envelope of a young, low mass, T Tauri star. One open question is whether the observed Brγ emission is caused by collisions or ionization by Galactic Centre stars. In the former case the emission is unrelated to the G objects being located in the vicinity of the black hole, which raises the question of why these objects have not been seen elsewhere.

In the third scenario, G1 and G2 are proposed to be binary merger products. The influence of the black hole will enhance the probability that binary systems merge through eccentricity oscillations due to the eccentric Kozai–Lidov (EKL) mechanism[28]. The merging process would inflate the outer layers of the merging binaries, which would produce an extended envelope of dust and gas around the merger product, hiding the central mass for an extended period of time. A few binary mergers are known in the Galaxy[49–51]. However, such mergers took place recently and were discovered because of the strong variability that probably characterizes the early stages of a merger. According to the merger hypothesis, the G objects are more likely to be in a much quieter long-term phase in which the merger has stabilized and is evolving slowly on a Kelvin–Helmholtz timescale. For this reason, it is not meaningful to compare the G objects to presently known mergers, especially because we still have scant quantitative knowledge of how a merger evolves.

The binary merger hypothesis could offer a mechanism to rejuvenate stars in the Galactic Centre, as in the case of blue stragglers[52,53] (but see[54]): some of the observed young stars orbiting closely around the central black hole (the S stars) could be the product of the merger of older stars. However, it is unclear whether this process can produce sufficiently massive stars to account for the S stars (typically 10–30 $M_\odot$[20]). The new star resulting from a merger can appear to be from a few Myr to several Gyr younger, depending on the merging circumstances[29].

Even if the G objects cannot account for the origin of the S stars, they are possibly connected to them. Here we have shown that the orbits of G3, G4, G5 and G6 have very different inclinations. This random distribution of the orbital planes very closely resembles the distribution of the orbits of S stars. If a stellar object is hidden inside a G object it must have a relatively small mass (less than a few solar masses), given the weakness of the continuum emission from these objects. In the central parsec, given the K′ detection limit[18], we can detect stars with masses down to ~$1.5 M_\odot$ (ref.[55]). However, we could detect low-mass binary systems that merge, producing a shell of dust and gas: gas would be ionized by the environmental radiation, while dust would be heated by both environmental radiation and the luminous energy emerging from the interior of the G object. The G objects could therefore offer a unique window on the low-mass, currently undetectable, part of the S-star cluster.

As a consistency check, we investigated whether the number of observed G objects is consistent with the expected number of binary mergers (see the following Methods section).

The EKL-induced binary merger hypothesis offers a compelling explanation for the origin of G objects that fits well with the three-body dynamics that are necessarily at play in a dense stellar environment, with the third body being a supermassive black hole. Moreover, a wide range of eccentricities is expected for such binary merger products[27], in agreement with what we observe.

**Binary fraction estimate**

To estimate the binary fraction from the current number of G objects, we assume that all observed G objects are binary merger products (indeed we expect a large fraction of binaries in the Galactic Centre based on the orbital configuration of the stellar disk[56]). We assume that all six of the G objects discussed here are relatively recent binary mergers, and that their progenitor binary systems were formed in the latest known star formation event 4–6 Myr ago. This assumption is supported by the fact that older binaries can only survive in the Galactic Centre if they are very tight, and therefore have a very low probability of merging[29]. They would consequently not contribute substantially to the observed population of G objects. We use binary merger rates[29] and the initial mass function[22]. Given the absence of continuum emission, we assume the G objects come from only the low-mass part of the population. Therefore, the binary fraction of low-mass stars is:

$$R = \frac{N_B}{N_m} \frac{1}{2} \qquad (1)$$

where $N_B$ is the number of binaries and $N_m$ the number of low-mass stars. We should expect about 10% of all binary systems to have merged within a few million years from a given star formation event in the Galactic Centre[29]. Also, 20%–25% of the initial binary population will have evaporated within the first few million years. So, given the observed number of G objects in the OSIRIS field of view ($N_G = 6$), the number of binaries present today is given by:

$$N_B = \frac{N_G}{0.1}(1 - 0.25 - 0.1) \qquad (2)$$

The initial mass function inferred by Lu et al.[22] is, $dN/dm = \xi m^{\wedge}(-\alpha)$ with $\alpha = 1.7$ and $\xi$ a normalization factor. Using this we can compute the number of low-mass stars ($1\,M_\odot < M < 10\,M_\odot$), $N_m$:

$$\int_{1M_\odot}^{10M_\odot} N_m = \xi \int_{1M_\odot}^{10M_\odot} m^{-\alpha} \qquad (3)$$

Given the number of stars we detect in the OSIRIS field of view ($N_M \approx 64$ stars with $M > 10 M_\odot$ is the average for 2018), we determine the normalization factor and estimate the number of low-mass stars from equation (3):

$$\int_{10M_\odot}^{30M_\odot} N_M = \xi \frac{M^{1-\alpha}}{1-\alpha} \Rightarrow \xi \sim 0.15 \Rightarrow N_m \sim 478.$$

From equations (1) and (2), it follows that the current binary fraction is $R \approx 5\%$. Locally in the Galaxy, the binary fraction of solar-type stars[30] is ~50%. Only about 20% of such binaries—which would be stable in the field—can be stable in the Galactic Centre[27], which leads to 10% solar-type star binaries in the Galactic Centre. 35% of these binaries have already evaporated after formation[29], resulting in a surviving binary fraction of 6%–7%, compatible with what we deduce from the observed abundance of G objects.